\documentclass[twocolumn]{aastex631}

\usepackage{xspace}
\usepackage[encapsulated]{CJK}  
\usepackage{gensymb}

\newcommand{\coto}{$^{12}{\rm CO}~(2-1)$\xspace}
\newcommand{\halpha}{H$\alpha$\xspace}
\newcommand{\hii}{\ion{H}{2}\xspace}
\newcommand{\hone}{H{\sc i}\xspace}
\newcommand{\hagasratio}{$I_{\rm H{\alpha}}/\Sigma_{\rm H{\sc I}+H_2}$\xspace}

\newcommand{\rpah}{$R_{\rm PAH}$\xspace}
\newcommand{\mjysr}{MJy~sr$^{-1}$\xspace}

\shorttitle{PAH Fraction and ISM in PHANGS-JWST}
\shortauthors{Chastenet, Sutter, Sandstrom et al.}

\begin{document}

\title{PHANGS-JWST First Results: Variations in PAH Fraction as a Function of ISM Phase and Metallicity}

\author[0000-0002-5235-5589]{J\'er\'emy Chastenet}
\affil{Sterrenkundig Observatorium, Ghent University, Krijgslaan 281-S9, 9000 Gent, Belgium}
\author[0000-0002-9183-8102]{Jessica Sutter}
\affiliation{Center for Astrophysics and Space Sciences, Department of Physics, University of California, San Diego\\9500 Gilman Drive, La Jolla, CA 92093, USA}
\author[0000-0002-4378-8534]{Karin Sandstrom}
\affiliation{Center for Astrophysics and Space Sciences, Department of Physics, University of California, San Diego\\9500 Gilman Drive, La Jolla, CA 92093, USA}

\author[0000-0002-2545-5752]{Francesco Belfiore}
\affiliation{INAF — Arcetri Astrophysical Observatory, Largo E. Fermi 5, I-50125, Florence, Italy}
\author[0000-0002-4755-118X]{Oleg V. Egorov}
\affiliation{Astronomisches Rechen-Institut, Zentrum f\"{u}r Astronomie der Universit\"{a}t Heidelberg, M\"{o}nchhofstra\ss e 12-14, 69120 Heidelberg, Germany}
\author[0000-0003-3917-6460]{Kirsten L. Larson}
\affiliation{AURA for the European Space Agency (ESA), Space Telescope Science Institute, 3700 San Martin Drive, Baltimore, MD 21218, USA}
\author[0000-0002-2545-1700]{Adam~K.~Leroy}
\affiliation{Department of Astronomy, The Ohio State University, 140 West 18th Avenue, Columbus, Ohio 43210, USA}
\author[0000-0001-9773-7479]{Daizhong Liu}
\affiliation{Max-Planck-Institut f\"ur Extraterrestrische Physik (MPE), Giessenbachstr. 1, D-85748 Garching, Germany}
\author[0000-0002-5204-2259]{Erik Rosolowsky}
\affiliation{Department of Physics, University of Alberta, Edmonton, Alberta, T6G 2E1, Canada}
\author[0000-0002-8528-7340]{David A. Thilker}
\affiliation{Department of Physics and Astronomy, The Johns Hopkins University, Baltimore, MD 21218, USA}
\author[0000-0002-7365-5791]{Elizabeth J. Watkins}
\affiliation{Astronomisches Rechen-Institut, Zentrum f\"{u}r Astronomie der Universit\"{a}t Heidelberg, M\"{o}nchhofstra\ss e 12-14, 69120 Heidelberg, Germany}
\author[0000-0002-0786-7307]{Thomas G. Williams}
\affiliation{Sub-department of Astrophysics, Department of Physics, University of Oxford, Keble Road, Oxford OX1 3RH, UK}
\affiliation{Max-Planck-Institut f\"{u}r Astronomie, K\"{o}nigstuhl 17, D-69117, Heidelberg, Germany}

\author[0000-0003-0410-4504]{Ashley~T.~Barnes}
\affiliation{Argelander-Institut f\"{u}r Astronomie, Universit\"{a}t Bonn, Auf dem H\"{u}gel 71, 53121, Bonn, Germany}
\author[0000-0003-0166-9745]{Frank Bigiel}
\affiliation{Argelander-Institut f\"ur Astronomie, Universit\"at Bonn, Auf dem H\"ugel 71, 53121 Bonn, Germany}
\author[0000-0003-0946-6176]{Médéric~Boquien}
\affiliation{Centro de Astronomía (CITEVA), Universidad de Antofagasta, Avenida Angamos 601, Antofagasta, Chile}
\author[0000-0002-5635-5180]{M\'elanie Chevance}
\affiliation{Institut f\"{u}r Theoretische Astrophysik, Zentrum f\"{u}r Astronomie der Universit\"{a}t Heidelberg,\\ Albert-Ueberle-Strasse 2, 69120 Heidelberg, Germany}
\affiliation{Cosmic Origins Of Life (COOL) Research DAO, coolresearch.io}
\author[0000-0003-2551-7148]{I-Da Chiang \begin{CJK*}{UTF8}{bkai}(江宜達)\end{CJK*}}%
\affiliation{Institute of Astronomy and Astrophysics, Academia Sinica, No. 1, Sec. 4, Roosevelt Road, Taipei 10617, Taiwan}
\author[0000-0002-5782-9093]{Daniel~A.~Dale}
\affiliation{Department of Physics and Astronomy, University of Wyoming, Laramie, WY 82071, USA}
\author[0000-0002-8804-0212]{J.~M.~Diederik~Kruijssen}
\affiliation{Cosmic Origins Of Life (COOL) Research DAO, coolresearch.io}
\author[0000-0002-6155-7166]{Eric Emsellem}
\affiliation{European Southern Observatory, Karl-Schwarzschild-Stra{\ss}e 2, 85748 Garching, Germany}
\affiliation{Univ Lyon, Univ Lyon1, ENS de Lyon, CNRS, Centre de Recherche Astrophysique de Lyon UMR5574, F-69230 Saint-Genis-Laval France}
\author[0000-0002-3247-5321]{Kathryn~Grasha}
\affiliation{Research School of Astronomy and Astrophysics, Australian National University, Canberra, ACT 2611, Australia}   
\affiliation{ARC Centre of Excellence for All Sky Astrophysics in 3 Dimensions (ASTRO 3D), Australia}   
\author[0000-0002-9768-0246]{Brent Groves}
\affiliation{International Centre for Radio Astronomy Research, University of Western Australia, 7 Fairway, Crawley, 6009 WA, Australia}
\author[0000-0002-8806-6308]{Hamid Hassani}
\affiliation{Department of Physics, University of Alberta, Edmonton, Alberta, T6G 2E1, Canada}
\author[0000-0002-9181-1161]{Annie~Hughes}
\affiliation{IRAP, Universit\'e de Toulouse, CNRS, CNES, UPS, (Toulouse), France} \author[0000-0001-6551-3091]{Kathryn Kreckel}
\affiliation{Astronomisches Rechen-Institut, Zentrum f\"{u}r Astronomie der Universit\"{a}t Heidelberg, M\"{o}nchhofstra\ss e 12-14, 69120 Heidelberg, Germany}
\author[0000-0002-6118-4048]{Sharon E. Meidt}
\affil{Sterrenkundig Observatorium, Ghent University, Krijgslaan 281-S9, 9000 Gent, Belgium}
\author[0000-0001-9719-4080]{Ryan J. Rickards Vaught}
\affiliation{Center for Astrophysics and Space Sciences, Department of Physics, University of California, San Diego\\9500 Gilman Drive, La Jolla, CA 92093, USA}
\author[0000-0002-5783-145X]{Amy Sardone}
\affiliation{Department of Astronomy, The Ohio State University, 140 West 18th Avenue, Columbus, OH 43210, USA}
\affiliation{Center for Cosmology and Astroparticle Physics, 191 West Woodruff Avenue, Columbus, OH 43210, USA}
\author[0000-0002-3933-7677]{Eva Schinnerer}
\affiliation{Max-Planck-Institut f\"{u}r Astronomie, K\"{o}nigstuhl 17, D-69117, Heidelberg, Germany}



\begin{abstract}
We present maps tracing the fraction of dust in the form of polycyclic aromatic hydrocarbons (PAHs) in IC~5332, NGC~628, NGC~1365, and NGC~7496 from JWST/MIRI observations.
We trace the PAH fraction by combining the F770W ($7.7~\mu$m) and F1130W ($11.3~\mu$m) filters to track ionized and neutral PAH emission, respectively, and comparing the PAH emission to F2100W which traces small, hot dust grains. We find average $R{\rm_{PAH} = (F770W+F1130W)/F2100W}$ values of 3.3, 4.7, 5.1, and 3.6 in IC~5332, NGC~628, NGC~1365, and NGC~7496, respectively.
We find that \hii~regions traced by MUSE \halpha show a systematically low PAH fraction. The PAH fraction remains relatively constant across other galactic environments, with slight variations. We use CO+\hone+\halpha to trace the interstellar gas phase and find that the PAH fraction decreases above a value of ${\rm I_{H\alpha}/\Sigma_{H{\sc I}+H_2}}~\sim~10^{37.5}~{\rm erg~s^{-1}~kpc^{-2}~(M_\odot~pc^{-2}})^{-1}$, in all four galaxies.  Radial profiles also show a decreasing PAH fraction with increasing radius, correlated with lower metallicity, in line with previous results showing a strong metallicity dependence to the PAH fraction. Our results suggest that the process of PAH destruction in ionized gas operates similarly across the four targets.
\end{abstract}



\section{Introduction}
\label{SecIntro}
The mid-infrared (mid-IR) emission features at 3.3, 6.2, 7.7, 8.6, 11.3, 12.6, and 17~$\mu$m are characteristic of the aromatic content of interstellar dust \citep[see reviews from][]{Draine2003Review, Tielens2008Review, Li2020Review}. The carriers of these features are often referred to as polycyclic aromatic hydrocarbons \citep[PAHs;][]{Allamandola1985}, and have been included as an extension of carbonaceous grains to small sizes in several physical dust models \citep[e.g.][]{DBP1990, Zubko2004, DL2007, Galliano2008}, or as an aromatic-rich mantle covering aliphatic grains \citep[e.g.][and reference therein]{THEMIS2017}. In the rest of this letter, we will refer to the carriers of mid-IR features as PAHs.

The (collective) brightness of these features with respect to a dust continuum emission can be used as a tracer of the fraction of dust in the form of PAHs.
The mass fraction of PAHs can be measured from fitting models to observed dust emission measured from mid- through far-IR broadband photometry \citep[e.g.][]{Draine2007, Galliano2008, Galliano2018, Chastenet2019, Nersesian2019, Aniano2020}.
One can also fit mid-IR spectra and derive more detailed information about the relative intensities of each feature, like the average charge and size of the PAH population \citep[][]{PAHFIT2007, Lai2020, Maragkoudakis2022}.
As very prominent features, the emission at 7.7 and 11.3~$\mu$m can be considered a satisfactory proxy to trace the total emission from PAHs in normal star-forming galaxies \citep[e.g.][]{PAHFIT2007, Lai2020, Draine2021}.

PAHs also play a key role in heating the ISM within photodissociation regions \citep[PDRs;][]{BakesTielens1994, WD2001, Tielens2008Review, Berne2009, Croxall2012, Wolfire2022}.  As a significant source of photo-ejected electrons, the abundance of PAHs greatly influences the photoelectric heating efficiency.  The close tie between PAHs and PDR heating has led to the suggested use of PAH emission as a tracer of the star-formation rate \citep[e.g.,][]{Peeters2004, Shipley2016}.  

The variations of the PAH fraction in the interstellar medium (ISM) of external galaxies helps us to understand their origin and evolution, as well as the mechanisms that regulate their formation and destruction. 
Several studies have found that the fraction of PAHs decreases in regions of ionized gas and/or because of hard radiation fields \citep[e.g.][]{Giard1994, Dong2011, Verstraete2011, Salgado2016, Chastenet2019, Rigopoulou2021}, as theory predicts \citep[][]{Siebenmorgen2004, Groves2008, Micelotta2010, Bocchio2012, Zhen2016}, and becomes very low in \hii~regions \citep[e.g.][]{Pety2005, Lebouteiller2007, Thilker2007, Compiegne2008}.
There is also evidence of a correlation of PAH features with the CO content \citep[see, e.g.,][]{LEROY1_PHANGSJWST}. Studying nearby galaxies, \citet[][]{Regan2006} found that the 8~$\mu$m (traced by \textit{Spitzer}/IRAC) and CO radial profiles are closely matched. Evidence of a close link between PAH and CO emission has also been observed in high-redshift galaxies on galactic scales \citep[e.g.][]{Pope2013, Cortzen2019}.

In addition to trends observed across different ISM environments, the abundance of PAHs has been shown to decrease in low metallicity galaxies \citep[e.g.][]{engelbracht2008,Draine2007,Sandstrom2012}.  This deficit in PAHs has multiple proposed causes, including the destruction of PAHs by hard radiation fields present in low-metallicity galaxies \citep{Madden2006, Gordon2008} or delayed PAH formation in AGB star atmospheres \citep{Galliano2008}.  These trends have important implications for future observations of PAH emission in high-redshift galaxies, making it essential to study PAH variation across a range of systems.  By further establishing how metallicity trends can lead to the reduction in PAH emission, we will be better prepared for using these small grains to assess ISM conditions across cosmic time.

The recent launch of JWST opens a new window for exploring this question.  The MIRI F770W and F1130W filters provide coverage of two of the most prominent PAH features at 7.7$\mu$m and 11.3$\mu$m, while the F2100W filter lends a useful comparison, tracing emission from larger dust grains \citep{DL2007}.  The unique ability of the MIRI instrument to map these PAH features at unprecedented resolution and sensitivity in galaxies outside of the Local Group allows us to greatly expand the range of ISM conditions in which measurements of the PAH fraction can be made and better determine how the local ISM conditions can influence the relative amount of PAHs present.

In this Letter, we investigate the variations of the PAH fraction traced by a combination of JWST/MIRI filters across the full disk of four nearby galaxies, IC~5332, NGC~628, NGC~1365, and NGC~7496.  By tracking the PAH fraction across the multi-phase ISM, we are able to determine how a range of ISM properties affect the relative abundance of PAHs with respect to large dust grains.  This will lay the ground work for more detailed studies of how PAH emission varies in specific conditions, such as those found around sites of active star formation.

\section{Data}
\label{SecData}
\subsection{JWST/MIRI data}
The data used in this paper are part of the PHANGS-JWST Treasury program \#2107 \citep[PI: J.C. Lee,][]{LEE_PHANGSJWST}.
We use MIRI \citep[][]{rieke2015} observations in the F770W, F1130W, and F2100W filters, from the latest reference files at the time of processing, as described by \citet[][]{LEE_PHANGSJWST} and \citet[][]{LEROY1_PHANGSJWST}.
The PHANGS-JWST team used the STScI Calibration pipeline 1.7.1 for NIRCam and 1.7.0 for MIRI, and Calibration Reference Data context number 0968 for both instruments.
Table~\ref{TabData} gives a few key details about the four targets of this Letter, which are used to construct $r/r_{25}$ maps.

We convolve all three filter maps to $1''$ resolution, which is larger than the full-width at half-maximum of the F2100W filter ($0.67''$).
The convolution kernels were computed using the theoretical PSFs of each filter provided by the \textsc{WebbPSF} \citep{WebbPSF}, and the method described in \citet[][]{Aniano2011}.
We choose to slightly degrade the data to increase the signal-to-noise (S/N) in the F2100W band. By doing so, we are also able to match the resolution of the CO and \halpha data (see below).

\subsection{Ancillary data}
We use \coto ``broad'' moment~0 maps and corresponding error maps from the PHANGS-ALMA survey combining ${\rm 12m+7m+Total~Power}$ \citep[][]{PHANGSALMA2021Pipeline, PHANGSALMA2021Survey}, to trace molecular gas, at $\sim 1''$ ($\sim 40 - 90$~pc in our sample) resolution. 
To convert CO intensity to molecular gas surface density, $\Sigma_{\rm H_2}$ in ${\rm M_\odot~pc^{-2}}$, we use a constant CO-to-H$_2$ conversion factor $\alpha_{\rm CO} = 4.35$~M$_\odot$~pc$^{-2}$~(K~km~s$^{-1})^{-1}$ as recommended for solar metallicity, star-forming galaxies \citep[][]{Bolatto2013}\footnote{With the assumption of a flat \hone distribution, the choice of $\alpha_{\rm CO}$ is not the dominant uncertainty.}, and \coto conversion factor and a $^{12}\mathrm{CO}(2-1)/^{12}\mathrm{CO}(1-0)$ line ratio $R_{21} = 0.65$ \citep[][]{denBrok2021, Leroy2022}.

We use \halpha maps from the PHANGS-MUSE survey \citep[][]{PHANGSMUSE2022Survey} to trace ionized gas. We use the `native' resolution maps, with an average $\sim 0.8''$ resolution, and a $0.2''$ pixel size.
To trace the ${\rm 12+log(O/H)}$ metallicity in our targets, we use the 2D maps by \citet[][]{Williams2022}.
The maps were created by interpolating \hii~regions-derived metallicity maps (with S-calibration), using a Gaussian Process Regression technique, based on  PHANGS-MUSE maps of \halpha intensity.
These maps have fixed physical-, and different angular resolutions \citep[][]{PHANGSMUSE2022Survey}, all lower than $1''$ except for NGC~1365 ($1.15''$).
We convolve all maps with a resolution lower than $1''$ to that value, to match the convolved MIRI maps.

We use \hone measurements from MeerKAT (C. Eibensteiner et al., in prep), at $15''$ resolution for NGC~7496, and from THINGS \citep[][]{Walter2008} for NGC~628 ($\sim 11''$ for the `natural' weighted map).
We convert the maps to have units of atomic gas mass surface density including helium assuming optically thin 21~cm emission, using the prescription from \citet[][see also \citeauthor{Walter2008} \citeyear{Walter2008}]{Leroy2012}.
We lack resolved 21~cm mapping for IC~5332 and NGC~1365. Based on the observed flatness of the atomic gas surface density over the inner parts of galaxy disks \citep[e.g.,][]{Schruba2011,  Bigiel2012, KennicuttReview, Wong2013}, we assume that the atomic gas has a flat distribution with ${\rm \Sigma_{H{\sc I}} = 8~M_\odot~pc^{-2}}$ (including helium).
The \hone data have the lowest resolution among the datasets for our sample. Since the \hone distribution is expected to be reasonably smooth across the disk of all our targets \citep[as a likely result of low resolution;][]{leroy2013}, we choose to work at the MIRI F2100W resolution to retain as much information about the distribution of the PAH emission within galaxies as possible.
We reproject these maps to the MIRI pixel grid, with pixel size $\sim 0.11''$.

\subsection{Noise properties and masks}
\label{subsec:mask}
We measure background standard deviation in NGC~7496 as it is the only one that offers enough pixels off target, at $1''$ resolution. \citet[][]{LEE_PHANGSJWST} give details of the reduction of the data, including noise. 
The background removal is done by anchoring the JWST data with \textit{Spitzer}, where the background in these sources was better estimated. This is done as few off-source pixels are available in most of the early targets.
We find that MIRI maps have similar noise, and we assume NGC~7496 estimates throughout the sample.
We remove pixels with signal-to-noise ${\rm S/N \leq 3}$ in all MIRI bands, with background noise values of 0.05, 0.05, and 0.1~\mjysr.
In NGC~1365 and NGC~7496, we mask pixels in the center due to IR brightness of the active galactic nuclei (AGN), which saturate the signal mostly in the F2100W band. This is done using the \textsc{WebbPSF} package \citep{WebbPSF}, and centering the F2100W PSF on the central coordinates of each target.
For this preliminary work, we remove additional conspicuous artifacts from the central AGN that remain after this analysis. 
These spikes are due to the central saturation, and although they are not flagged as `bad pixels' by data reduction, they are an obvious artifact. We mask these by hand to ensure they do not bias our initial results. They are shown in gray scale in Figure~\ref{FigAllMaps}, but are not used in the analysis.

\begin{table*}[]
    \centering
    \begin{tabular}{|lcccccc|cc|}
    \hline
        \textbf{Target} & \textbf{R. A.} & \textbf{Dec} & \textbf{Distance} [Mpc] & \textbf{P.A.} [\degree] & \textbf{\textit{i}} [\degree] & ${\rm r_{25}}$ [$'$] & $\langle R_{\rm PAH} \rangle $ & ${\rm 16^{th}-84^{th}}$ perc.\\
        \hline
        IC~5332 & $23:34:27.488$ & $-36:06:3.89$ & 9.01 & 74.4 & 26.9 & 3.03 & 3.3 & $1.8 - 4.8$ \\
        NGC~628 & $01:36:41.745$ & $+15:47:1.11$ & 9.84 & 20.7 & 8.9 & 4.94 & 4.7 & $3.8 - 5.6$ \\
        NGC~1365 & $03:33:36.458$ & $-36:08:26.37$ & 19.57 & 201.1 & 55.4 & 6.01 & 5.1 & $3.8 - 6.3$ \\
        NGC~7496 & $23:09:47.288$ & $-43:25:40.28$ & 18.72 & 193.7 & 35.9 & 1.67 & 3.6 & $1.8 - 5.3$ \\
        \hline
    \end{tabular}
    \caption{Right ascension (R.~A.) and declination (Dec) coordinates (J2000), the ${\rm r_{25}}$ radius, in arc-minutes, used in this Letter for our four targets, from the HyperLeda database \citep[][]{Makarov2014}. Distances are from \citet[][]{PHANGS_DISTANCE}. Position angles and inclinations are from \citet[][]{PHANGSALMA2021Survey}. Additional information can be found in Table~1 of the survey paper \citep[][]{LEE_PHANGSJWST}. The second part of the Table shows the mean and ${\rm 16^{th}-84^{th}}$ percentiles of \rpah.}
    \label{TabData}
\end{table*}

\section{PAH Fraction}
\label{SecAbundance}
In \citet{Draine2021}, the authors found that the luminosity of the 7.7~$\mu$m feature\footnote{In this model, the 7.7~$\mu$m feature luminosity is determined by integrating between two set ``clip'' points, $\lambda=6.9$ and $\lambda=9.7$.  This is not exactly equal to fluxes observed using the F770W filter, which spans $\lambda=6.6-8.6$, but is similar.} normalized to the total IR luminosity can approximate the PAH fraction in all but the most extreme cases \citep{DL2007, Draine2021}.
Here, we use the JWST/MIRI $R{\rm_{PAH} = (F770W+F1130W)/F2100W}$ ratio as a proxy for the PAH fraction.
Over a broad range of PAH size (in terms of number of carbon atoms), and assuming a Galactic interstellar radiation field \citep[][at 10~kpc]{MMP1983}, the 7.7~$\mu$m feature is more representative of the ionized PAH population, and the 11.3~$\mu$m feature of the neutral population \citep[e.g.,][]{Rapacioli2005, Draine2021}.
We assume the F2100W filter is free from contribution from PAHs and is instead dominated by small, hot dust grains, as seen in the \textit{Spitzer}/IRS spectra of local galaxies observed by the SINGS project \citep[][see also \citeauthor{DL2007} \citeyear{DL2007}, \citeauthor{Dale2009} \citeyear{Dale2009}]{PAHFIT2007}.  This assumption is validated by a clear difference in behavior between the F2100W and PAH--dominated band emission, demonstrated in \citet[][]{LEROY1_PHANGSJWST}. Normalizing the emission at 7.7 and 11.3~$\mu$m by the observed flux at 21~$\mu$m help us focus on the PAH-only fraction.  Previous studies of the PAH population in nearby galaxies completed with \textit{Spitzer} have shown that the ratio of 8~$\mu$m to 24~$\mu$m is a good tracer of the PAH fraction \citep[see e.g.][]{PAHFIT2007, Marble2010, Croxall2012}.  In addition, models of PAH and dust emission have shown 7.7~$\mu$m-to-Total~IR luminosity is a good tracer the PAH mass fraction \citep[$q_{\rm{PAH}}$;][]{DL2007}.  
Using the MIRI filters, we can improve on this by using the bands centered on the PAH features (F770W and F1130W) in place of the 8~$\mu$m data while the F2100W replaces the 24~$\mu$m continuum tracer.  As both the PAH features and the 21~$\mu$m continuum are from stochastic heating, using these fluxes to determine the PAH fraction removes any significant dependence on the radiation field \citep[except in extreme situations, see][their Figure 13]{DL2007}.

\subsection{Qualitative description}
\begin{figure*}
    \centering
    \includegraphics[width=\textwidth, clip, trim={2cm 2cm 2cm 0.5cm}]{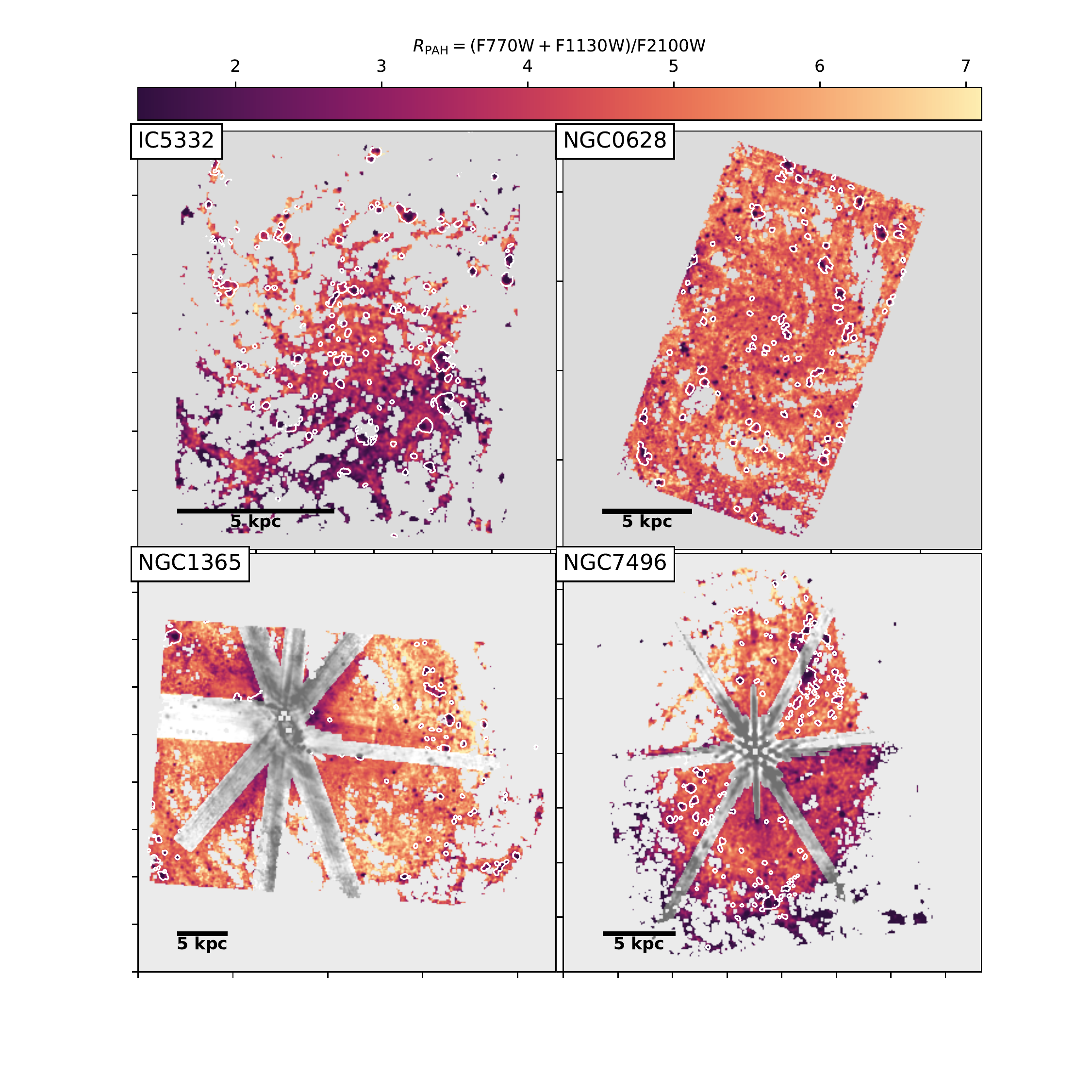}
    \caption{Maps of \rpah in IC~5332 (top left), NGC~628 (top right), NGC~1365 (bottom left), and NGC~7496 (bottom right). We mask the pixels with a ${\rm S/N < 3}$ in all bands (gray uniform background).
    We also mask the central pixels in NGC~1365 and NGC~7496 which are saturated, using instrument PSFs, and perform a by-hand additional masking to remove conspicuous saturation artifacts (shown in gray scale, not included in the analysis).
    We plot contours for a few of the brightest \hii~regions. They are clearly visible as depressions (darker colors) in \rpah, especially in NGC~628 and NGC~7496.}
    \label{FigAllMaps}
\end{figure*}

Figure~\ref{FigAllMaps} shows \rpah in the four PHANGS-JWST early targets. 
The white contours show the brightest \hii~regions \citep[][]{GROVES_HIICAT, PHANGSMUSE2022HII}.
In all cases, it appears that \rpah shows clear depressions within \hii~regions, though NGC~628 and NGC~7496 show this contrast most clearly.

In the right section of Table~\ref{TabData}, we show the mean of the ratio, $\langle R_{\rm PAH} \rangle$, and associated $16^{\rm th}-84^{\rm th}$ percentile range. 
NGC~1365 shows the highest average for \rpah, and IC~5332 the lowest. All $16^{\rm th}-84^{\rm th}$ ranges agree reasonably well, and NGC~7496 shows the broadest range.
The top left panel of Figure~\ref{FigMediansPanel} shows the radial profiles of \rpah, in bins of $r/r_{25}$. In all panels, the error bars are 3 times the standard error of the mean (SEM). These errors are small because of the number of pixels in each bin (the standard deviation is much larger).
Note that not all galaxies have observations extending to the same $r/r_{25}$. 

All galaxies show an overall decreasing trend with radius. 
There is however, a slightly different trend in NGC~1365 and NGC~7496, which both show an increase in abundance ratio at small radii before exhibiting a steady decrease. Both are Seyfert galaxies \citep[NGC~1365: 1.8, NGC~7496: 2.0; e.g.,][]{GarciaBernete2022} hosting an AGN as well as central bars that feed high density regions at their centers. Even though we mask the region around the AGN, this increasing trend at small radii could be due to the influence of the central AGN on the PAH population. 
The definitive impact of AGNs on the PAH population is not yet completely clear.
For example, \citet{GarciaBernete2022} found differences in the relative strengths of different PAH features in AGN host galaxies and star-forming galaxies at kpc scales.  
However, \citet[][]{Lai2022} recently found relatively small variations in PAH size and ionization and a decrease in PAH emission only in the \textit{direct} line-of-sight of the AGN, using JWST observations.
Similarly, \citet[][]{Viaene2020} found that the influence of the AGN is only relevant close to the nucleus.
Additionally, there is evidence for PAH molecules surviving the harsh environment surrounding AGNs \citep[e.g.,][]{Jensen2017, Alonso-Herrero2014, GarciaBernete2022}.
Although we observe a decreasing trend in these two galaxies towards the center, it is as of yet uncertain how the presence of an AGN may be related to this decrease. Additional work will be required to clearly understand the scales on which AGN impact the global fraction of PAHs.

We also show the profile of \rpah with ${\rm 12+log(O/H)}$ metallicity, in the top right panel of Figure~\ref{FigMediansPanel}.
Global trends of the PAH fraction with metallicity have been studied on integrated scales in several works \citep[][]{Draine2007, RemyRuyer2015, Chastenet2019, Galliano2021}. Here, we see that this trend is similar on small resolved scales ($\sim$~tens of parsecs). 
The turning point in the metallicity value, where the abundance ratio starts to decrease with increasing metallicity, is similar to the down-turn at small $r/r_{25}$ as these are primarily radial metallicity gradients.

\subsection{Variation of PAH fraction with ISM environment}
\begin{figure*}
    \centering
    \includegraphics[width=\textwidth]{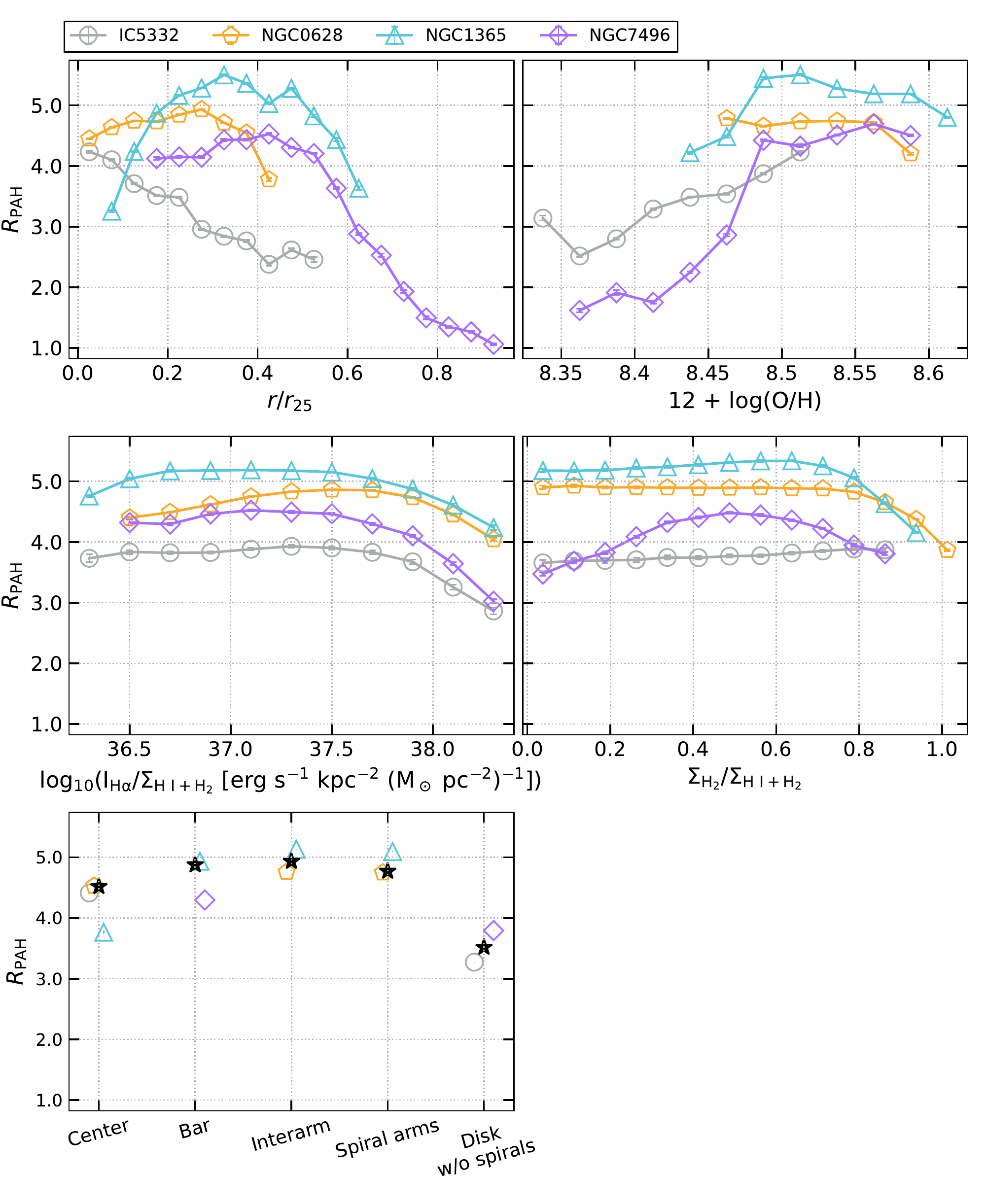}
    \caption{Running medians of \rpah, as a function of 
    (\textit{top left:}) $r/{\rm r_{25}}$; 
    (\textit{top right:}) ${\rm 12+log(O/H)}$ using metallicity maps from \citet[][]{Williams2022};  
    (\textit{middle left:}) \hagasratio in units of ${\rm  erg~s^{-1}~kpc^{-2}~(M_\odot~pc^{-2}})^{-1}$; 
    (\textit{middle right:}) the fraction of molecular gas; 
    (\textit{bottom left:}) the environmental masks from \citet[][]{Querejeta2021}, with a black star symbol showing the median for all pixels within each category.
    The error-bars show $3 \times$~standard error of the mean in each bin (except for the bottom left panel, only 1 SEM).
    Note that the middle panels involve the flat \hone distribution assumption in IC~5332 and NGC~1365, which may shift the curves horizontally.}
    \label{FigMediansPanel}
\end{figure*}

In Figure~\ref{FigMediansPanel}, middle and bottom rows, we present the variation of \rpah as a function of the ISM environment.
We combine the \hone and H$_2$ maps to estimate the total gas surface density, and use the ratio of \halpha intensity to total gas to follow the variations of the ISM conditions in each target (in units of ${\rm erg~s^{-1}~kpc^{-2}~(M_\odot~pc^{-2})^{-1}}$). 
As we expect PAH emission to arise from a range of ISM phases, the ratio of \halpha intensity to total gas (\hagasratio) is used as a proxy for environments dominated by the ionized phase and can provide a clear indication of what ISM environments the PAH emission is coming from.  For example, high values of \hagasratio are indicative of regions dominated by ionized gas, while low values of \hagasratio suggest the dominance of neutral gas.  
This comparison works well because neutral and ionized gas are decorrelated on small spatial scales and the clearing of neutral gas due to rapid ionizing feedback \citep[$<5~\mathrm{Myr}$,][]{Kruijssen2019, Chevance2020, Chevance2022, Kim2022}.
By comparing \rpah to \hagasratio we can better understand what ISM conditions, i.e. what fraction of the line of sight is ionized gas free of PAHs, could lead to an observed dearth of PAH emission. This is further investigated within \hii~regions in \citet[][]{EGOROV_PHANGSJWST}.

The middle left panel of Figure~\ref{FigMediansPanel} shows the variations of \rpah as a function of \hagasratio.
The abundance of PAHs appears to stay rather flat until a threshold in \hagasratio, where it decreases steeply. This is expected from the destruction of PAHs in harsh environments traced by high-intensity \halpha \citep[e.g.,][]{Groves2008, Micelotta2010, Bocchio2012, EGOROV_PHANGSJWST}, and has been observed in Galactic PDRs \citep[][]{Pety2005, Compiegne2008}.
Interestingly, it appears that all galaxies share similar thresholds in \hagasratio at which the PAH fraction starts to decrease. These inflection points seem to be around $\sim 37.5~{\rm  erg~s^{-1}~kpc^{-2}~(M_\odot~pc^{-2}})^{-1}$ for all targets. However, it should be pointed out that there are (currently) no similar resolution \hone data for IC~5332 and NGC~1365, and therefore a universal threshold across all environments is left for future studies.
Considering \hagasratio traces the fraction of ionized gas per unit of total gas, this common threshold could be a limit at which the amount of radiation producing the intense H$\alpha$ emission is able to destroy the PAHs through sputtering, overcoming shielding from molecular gas and reducing the observed PAH fraction. 

We can also see that there is an offset in PAH fraction, on average, between each galaxy (see also Table~\ref{TabData}). 
Overall, this offset nicely tracks the global galaxy metallicity gradients \citep[][]{Kreckel2019, PHANGSMUSE2022HII, GROVES_HIICAT}, in the lower \halpha intensity regions.
As we move towards higher \hagasratio values, the offset gets minimized. This relates to results seen by \citet[][]{EGOROV_PHANGSJWST}, where no metallicity trend is found with \rpah within \hii~regions.
This suggests that the offset in PAH fraction between the four galaxies may be driven by a difference in the PAH population in the diffuse or neutral ISM set by the average metallicity of the galaxy.  
This general offset also follows trends observed in previous works that found that the PAH fraction correlates positively with metallicity, in nearby galaxies \citep[][]{Draine2007, RemyRuyer2015, Chastenet2019, Galliano2021}, although it should be noted that the sample included in this work covers a small range of metallicity. 

\begin{figure*}
    \centering
    \includegraphics[width=\textwidth, clip, trim={1cm 1cm 2cm 0.5cm}]{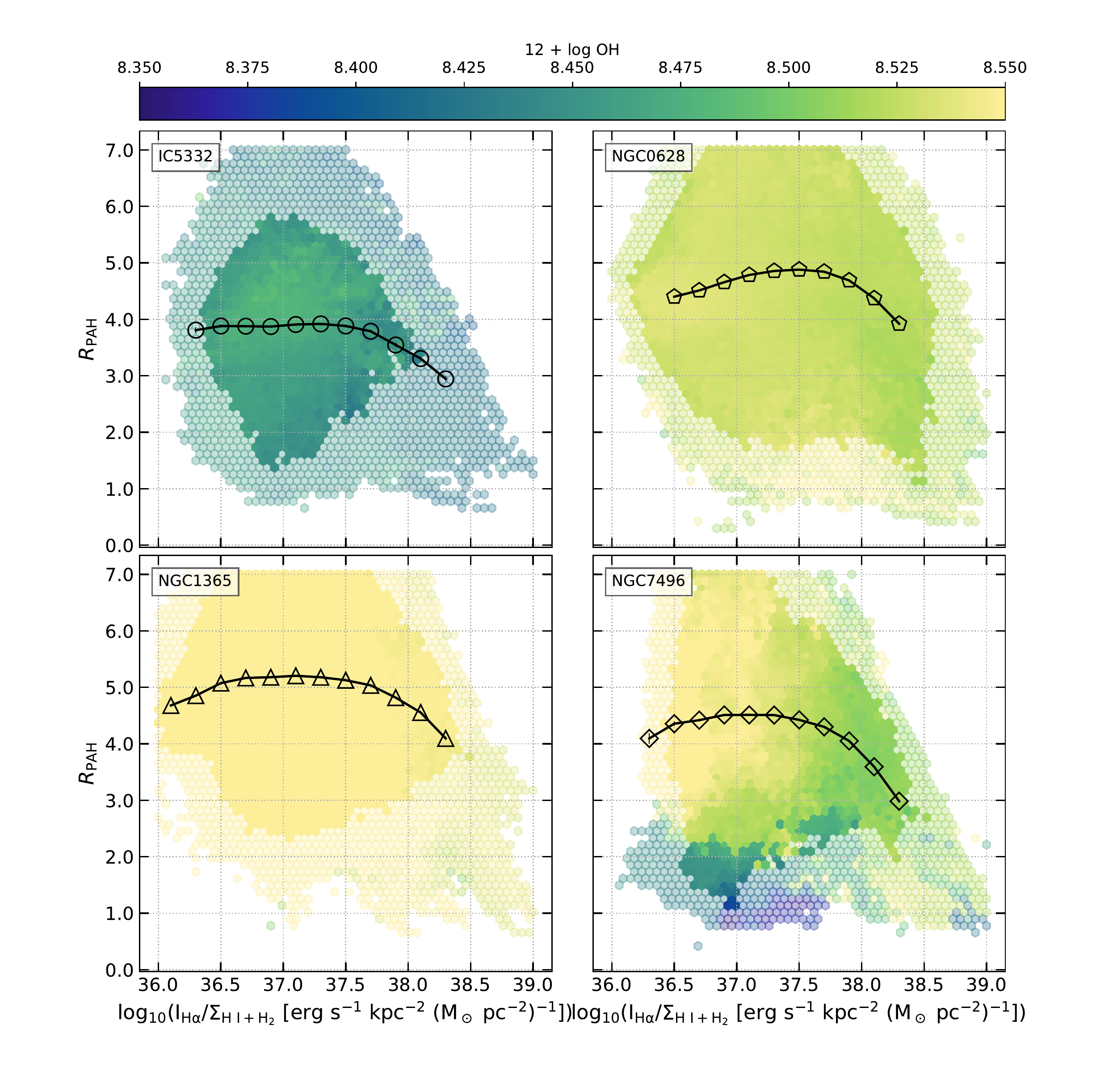}
    \caption{2D histograms of the \hagasratio and the \rpah, color-coded by metallicity, using the gradient from PHANGS-MUSE. 
    The more transparent colors indicate bins with at least 10 hits, while the solid colors with at least 100 hits per bin.}
    \label{FigHaGas_Ratio2D}
\end{figure*}

Figure~\ref{FigHaGas_Ratio2D} shows the same variations, with metallicity information. We show the 2D-histograms of \rpah as a function of \hagasratio, color-coded by the median ${\rm 12+log(O/H)}$ metallicity in each bin. 
There is a visible color difference between each galaxy, but no clear gradient.
This implies that while the average metallicity of each galaxy seems to influence the PAH fraction, the moderate local metallicity variations are not having a large effect on \rpah.

In the middle right panel of Figure~\ref{FigMediansPanel}, we show the variations of \rpah as a function of the fraction of molecular gas to total cold gas fraction, as traced by CO. 
Behaviors vary between each galaxy, showing either a similar profile to that of \rpah with \hagasratio \citep[NGC~1365, NGC~628, reflecting the similarity in spatial distribution of CO and \halpha emission at these scales][]{Schinnerer2019}, a rather flat trend (IC~5332, which has little CO), or a $\sim 20\%$ increase to a maximum value, followed a decrease in \rpah (NGC~7496).
This could suggest that the abundance of PAHs is not particularly sensitive to the molecular fraction, compared to the ionized gas content.
\citet[][]{CHASTENET2_PHANGSJWST} found that the average grain size of the global PAH population \citep[as traced by the 3.3/7.7~$\mu$m ratio; e.g.,][]{Maragkoudakis2020, Draine2021, Rigopoulou2021} is more sensitive to the fraction of molecular gas.

\begin{figure*}
    \centering
    \includegraphics[width=\textwidth, clip, trim={2cm 1cm 1cm 0}]{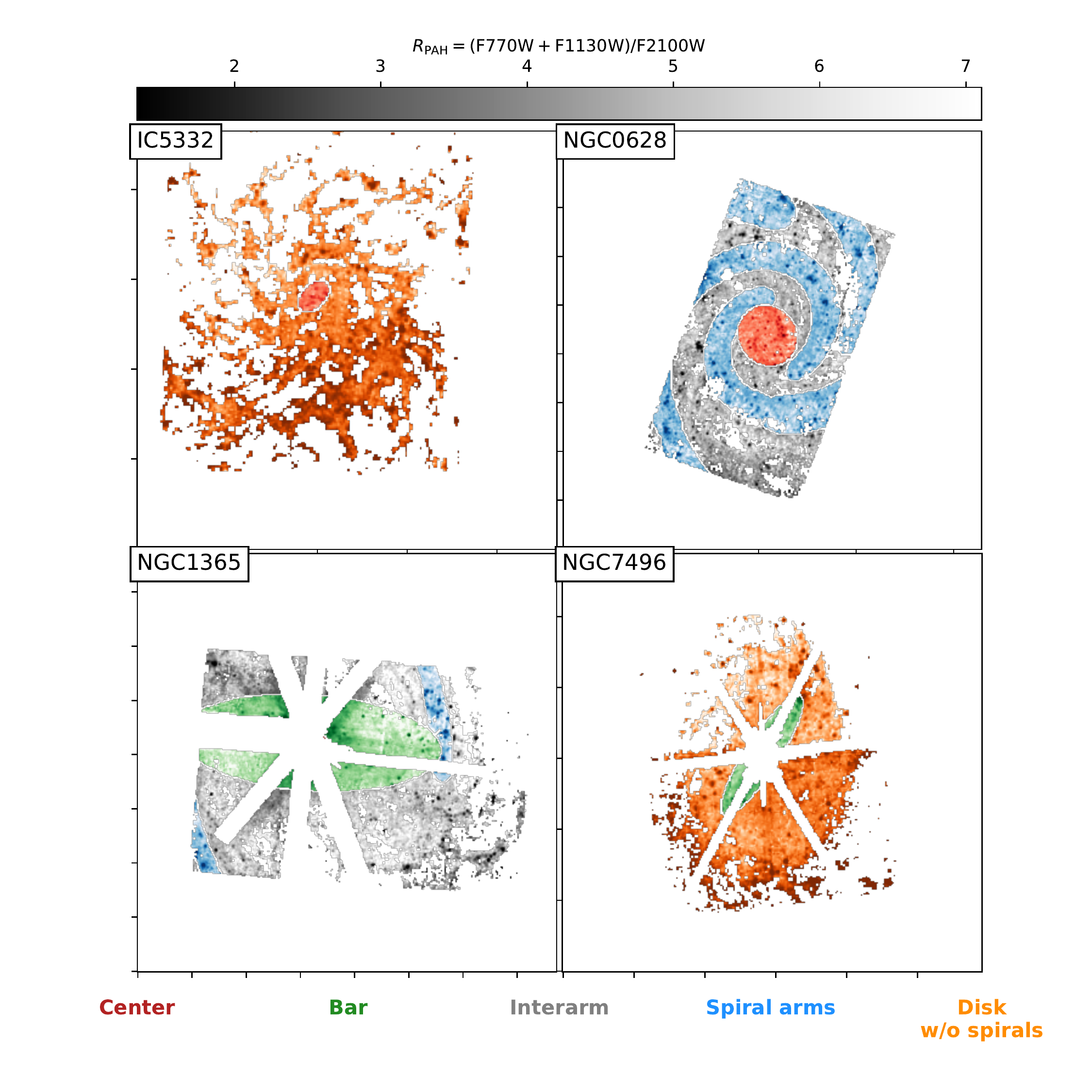}
    \caption{\rpah maps colored by the environmental masks from \citet[][]{Querejeta2021}: center in red, bar in green, interarms in gray, spiral arms in blue and disk in orange. We use this separation to measure the median \rpah in each phase individually, and collectively, in Figure~\ref{FigMediansPanel}.}
    \label{FigEnvMasks}
\end{figure*}

In the bottom left panel of Figure~\ref{FigMediansPanel}, we plot the median values of \rpah in different galactic environments,  identified by \citet[][]{Querejeta2021}. We use their spatial mask to separate pixels in 5 different categories (see their Table~1). 
Figure~\ref{FigEnvMasks} shows the masks projected to the \rpah maps, for a visual representation of the different environments.
In this panel, it appears that there is no striking differences between environments, again showing minimal variations of a few tens of percent, for individual galaxies.
We also show the median and associated standard error of the mean (SEM) for \emph{all} pixels falling into each category, with black symbols. Here, we can see that \rpah is the highest in the bar, and interarm regions, with lower values in the spiral arms, followed by the center and finally in the disk. Although this approach is limited by the number of targets at this stage, it shows promising results. The higher fraction of PAHs in the interarms tracks with a less harsh environment due to star formation, and a possibly more \hone dominated ISM. Adding more targets to this approach will provide a generic view of the variation of the PAH fraction in the different phases of nearby galaxies.

Future work will improve on this work by more finely binning the data. For example, it will be interesting to investigate the sensitivity of each parameter to the S/N measured in the MIRI data. 
It will also be possible to test a Vorono\"i binning, to check whether the trends seen in Figure~\ref{FigMediansPanel} would be significantly pulled down by taking into account more low-S/N pixels.

\section{Conclusions}
With the advent of the JWST, we can now probe individual mid-IR emission features on spatial scales never achieved before outside the Local Group. In this letter, we have used a combination of the JWST/MIRI F770W, F1130W, and F2100W filters to trace the abundance of PAHs relative to small dust grains in four nearby galaxies, IC~5332, NGC~628, NGC~1365, and NGC~7496, as part of the Treasury GO program PHANGS-JWST \#2107. 

We present maps of \rpah$\equiv\mathrm{(F770W + F1130W)/F2100W}$ in these first four targets.  This ratio traces the relative fraction of PAHs (the F770W and F1130W bands) to small dust grains (from the F2100W band).  The ratio \rpah decreases in \hii~regions, showing that the PAH fraction drops there, which is further discussed in \citet[][]{EGOROV_PHANGSJWST}.
We track the variations of the abundance ratio as a function of the ISM content as traced by CO, \halpha, \hone, and metallicity measurements.
We find that \rpah as a function of ionized gas fraction (traced by \hagasratio, Figure~\ref{FigMediansPanel}, bottom left panel) shows a similar trend in all the targets: a rather flat distribution up to a value of ${\rm I_{H\alpha}/\Sigma_{H{\sc I}+H_2}}~\sim~10^{37.5}~{\rm erg~s^{-1}~kpc^{-2}~(M_\odot~pc^{-2}})^{-1}$ for all galaxies, at which the abundance ratio systematically decreases.
The variations with the fraction of molecular gas (Figure~\ref{FigMediansPanel}, bottom right panel) are rather small.  This work sets the stage for future research to refine how the local environment influences the relative PAH fraction.  As JWST data reduction methods are improved and the sample of galaxies with this coverage expands, the conditions in which PAHs are found will be better established.  This early study provides insights into how global metallicity and ISM environment can effect the relative PAH population, and shows the improvements that JWST observations bring to determining the answers to these questions.

\section*{Acknowledgments}
We thank the anonymous referee for their careful reading and comments that helped improve the clarity of the paper.
This work was carried out as part of the PHANGS collaboration, associated with JWST program 2107. This work is based on observations made with the NASA/ESA/CSA JWST. 
Some/all of the data presented in this paper were obtained from the Mikulski Archive for Space Telescopes (MAST) at the Space Telescope Science Institute, which is operated by the Association of Universities for Research in Astronomy, Inc., under NASA contract NAS 5-03127.
The specific observations analyzed can be accessed via \dataset[10.17909/9bdf-jn24]{http://dx.doi.org/10.17909/9bdf-jn24}. 
Based on observations collected at the European Southern Observatory under ESO programmes 094.C-0623 (PI: Kreckel), 095.C-0473,  098.C-0484 (PI: Blanc), 1100.B-0651 (PHANGS-MUSE; PI: Schinnerer), as well as 094.B-0321 (MAGNUM; PI: Marconi), 099.B-0242, 0100.B-0116, 098.B-0551 (MAD; PI: Carollo) and 097.B-0640 (TIMER; PI: Gadotti). 
This paper makes use of the following ALMA data: \linebreak
ADS/JAO.ALMA\#2012.1.00650.S, \linebreak 
ADS/JAO.ALMA\#2013.1.01161.S, \linebreak 
ADS/JAO.ALMA\#2015.1.00925.S, \linebreak 
ADS/JAO.ALMA\#2015.1.00956.S, \linebreak 
ADS/JAO.ALMA\#2017.1.00392.S, \linebreak 
ADS/JAO.ALMA\#2017.1.00766.S, \linebreak 
ADS/JAO.ALMA\#2017.1.00886.L, \linebreak 
ADS/JAO.ALMA\#2018.1.01651.S. \linebreak 
ADS/JAO.ALMA\#2018.A.00062.S. \linebreak 
ALMA is a partnership of ESO (representing its member states), NSF (USA) and NINS (Japan), together with NRC (Canada), MOST and ASIAA (Taiwan), and KASI (Republic of Korea), in cooperation with the Republic of Chile. The Joint ALMA Observatory is operated by ESO, AUI/NRAO and NAOJ.

JC acknowledges support from ERC starting grant \#851622 DustOrigin.
EJW acknowledges funding from the Deutsche Forschungsgemeinschaft (DFG, German Research Foundation) -- Project-ID 138713538 -- SFB 881 (``The Milky Way System'', subproject P1). 
MC gratefully acknowledges funding from the DFG through an Emmy Noether Research Group (grant number CH2137/1-1).
COOL Research DAO is a Decentralized Autonomous Organization supporting research in astrophysics aimed at uncovering our cosmic origins.
JMDK gratefully acknowledges funding from the European Research Council (ERC) under the European Union's Horizon 2020 research and innovation programme via the ERC Starting Grant MUSTANG (grant agreement number 714907).
TGW and ES acknowledge funding from the European Research Council (ERC) under the European Union’s Horizon 2020 research and innovation programme (grant agreement No. 694343).
MB acknowledges support from FONDECYT regular grant 1211000 and by the ANID BASAL project FB210003.
IC thanks the National Science and Technology Council for support through grants 108-2112-M-001-007-MY3 and 111-2112-M-001-038-MY3, and the Academia Sinica for Investigator Award AS-IA-109-M02.
KK, OE gratefully acknowledge funding from the Deutsche Forschungsgemeinschaft (DFG, German Research Foundation) in the form of an Emmy Noether Research Group (grant number KR4598/2-1, PI Kreckel). 
FB would like to acknowledge funding from the European Research Council (ERC) under the European Union’s Horizon 2020 research and innovation programme (grant agreement No.726384/Empire).
ER and HH acknowledge the support of the Natural Sciences and Engineering Research Council of Canada (NSERC), funding reference number RGPIN-2022-03499.
KG is supported by the Australian Research Council through the Discovery Early Career Researcher Award (DECRA) Fellowship DE220100766 funded by the Australian Government. 
KG is supported by the Australian Research Council Centre of Excellence for All Sky Astrophysics in 3 Dimensions (ASTRO~3D), through project number CE170100013. 
AS is supported by an NSF Astronomy and Astrophysics Postdoctoral Fellowship under award AST-1903834.
AKL gratefully acknowledges support by grants 1653300 and 2205628 from the National Science Foundation, by award JWST-GO-02107.009-A, and by a Humboldt Research Award from the Alexander von Humboldt Foundation.


\facilities{JWST (MIRI), MUSE, ALMA}

\bibliography{main,phangsjwst}{}
\bibliographystyle{aasjournal}



\end{document}